# Detection of Solar and Lunar Tidal Forces via Non-resonant Oscillations of A Pendulum

Timir Datta[1], Ming Yin[2], Mike Wescott[1], Yeuncheol Jeong[1], Huaizhou Zhang[1]
1: Physics & Astronomy Department
University of South Carolina, Columbia, SC 29208, USA
2: Physics and Engineering, Benedict College, Columbia, SC 29204, USA
And
George Voulgaris
Department of Earth and Ocean Sciences
University of South Carolina, Columbia, SC 29208, USA

Overview:

Pursuant of gravitational interactions between source and proof masses in the laboratory we have analyzed time deflection data of a pendulum bob. In the FFT power spectrum a broad (~ $\mu$Hz) structure consistent with non-resonant response due to forced acceleration (~$10^{-7}$m/s$^2$) was observed. By comparative harmonic analyses of the 30 min low pass Butterworth filtered data and the horizontal components of the theoretical tidal forces we conclude that the observed diurnal and semi-diurnal oscillations are due to solar and lunar tides. In high precision metrology, tidal modulations in the vertical free–fall acceleration g are routinely noted. Our results presented here clearly indicate that under the appropriate conditions the horizontal forces are also present and have to be accounted for. To the best of our knowledge this is the first reported detection of the horizontal tides with a pendulum.

With currently available Laser sources, charge coupled device (CCD) detectors, high throughput interfacing and digital computers with large storage capacities it is possible to effortlessly acquire many terabits of quantitative information over an extended length of time. Long period investigations are essential when precise information regarding frequency, period or stability of oscillation is needed.

We have been engaged in a study of gravitational interactions [1-3] between a proof mass (a torsion pendulum bob) and source masses inside a laboratory in the basement of a large building situated on the campus of an urban university. For this purpose, it is imperative that under a given fixed distribution of the source masses the time displacement behavior of the pendulum bob be accurately reproduced over many sets of experiments [1-4]. In this article we will describe a non-resonant periodic response of the proof mass.

Pendulum was selected because Simple Harmonic Oscillators (SHO) are the instruments of choice in force metrology [5,6]. Pendula are the first realizations of SHO. Harmonic detectors are extremely versatile with applications ranging from atomic force microscopy to the macroscopic gravity meters. This preference for SHO reflects the famed isochronous property – that is the frequency of oscillation is insensitive to the details of apparatus design and construction as well as the amplitude of vibration. The necessary conditions for SHO are that the systems have stable equilibrium and perturbations away from equilibrium produce linear restoring response [7-11]. A mechanical oscillator comprises of an inertial (massive) object [12,13] or the bob subjected to a mechanism to produce "Hook's law" force. Remarkably, isochronous behavior can be retained to any desired level of constancy provided some minimal condition such as linearity is maintained, also frictional effects can be accounted for by the introduction a the appropriate dissipative function.

Our pendulum is the central component of an apparatus designed to measure the force of gravitational interactions between the proof mass (pendulum bob) and the field or source masses for an experimental determination of Newton's constant [1-4]. The bobs were uniform, metal cylinders of aluminum, copper and stainless steel machined out of commercially procured single stocks of solid bars. The geometry varied over a restricted range of length and diameter consistent with a maximum spread of aspect ratios plus the space available in the vacuum chamber, the cylindrical shape was chosen to achieve the best dimensional tolerance for the given machining cost. The mass of the bob is in the ~0.1 - 1 kg range. During a run the chosen bob was freely suspended (with the long cylinder axis horizontal) in a 2.5 m tall stainless steel vacuum chamber by a single monofilament fiber ~ 50 μm in diameter and ~ 2m in length. A diagram of the apparatus is shown in figure 1.

The displacement of the center of mass of the pendulum relative to a fixed coordinate system was optically amplified with a HeNe laser beam and a CCD photon detector. The position coordinates of the center of intensity (COI) of the laser beam, along a direction at 21.5 degrees North was analyzed at a rate of 30Hz with a 100nm resolution and digitally recorded along with time. The apparatus was located at a subbasement laboratory in an urban university campus, with coordinates 33.9992N 81.0264W and an elevation of 92.0m above datum.

The suspension provides several essential functions, first is the obvious holding of the bob, secondly it is the elastic element responsible for the low frequency (~ mHz) azimuthal oscillations. Furthermore, due to internal friction the fiber produces Raleigh damping as well as function as the dissipative dashpot for filtering the high frequency noise. The system was mechanically isolated from the surroundings by several layers of additional passive vibration filters. Nevertheless, it was ceaselessly excited by remnant tremors from its surroundings that were not filtered out. The bob was in thermal equilibrium and hence was observed to be in constant motion, performing oscillations about the fiber axis, point of suspension as well as about body-axes passing through its center of mass.

The principle mode was the torsional oscillations in the azimuthal angle, $\phi(t)$, executed about the vertical axis. Next is simple Galilean pendulation about the point of suspension i.e., periodic variations of the apex angle $\theta(t)$. The eponymous rigid up-and-down bobbing motion z(t) of the bob and rocking about the center of mass of the cylinder $\Omega(t)$. These four vibrations are shown in figure

2. Consistent with the imperfect conical symmetry due with the attitude of the cylindrical bob, the notoriously fastidious Foucault mode is greatly suppressed.

However even a perfectly rigid solid cylinder attached to an elastic fiber which is suspended from an infinitely rigid support is in reality a highly coupled system. Where due the constraints of fiber attachment and suspension, changes in one degree of freedom affects changes in another, for example the tendency to rock or see-sawing of the cylinder axis about a horizontal diameter passing through the center of mass (small red circle) causes changes in the apex angle ($\theta^*$). Even the Galilean mode is not pure realistically it is more like a double pendulum. These couplings force the center of mass to be displaced vertically as shown in figure 3, and because of the large mass changes the total energy significantly hence are of higher frequency.

Changes in $\theta(t)$ produced horizontal displacements of the center of mass of the bob is the principle topic of this article. As will be shown below these forced periodic horizontal displacements are due to tidal forces of astronomical origin.

The raw COI data contained noise due to vibrations of the building structure, local geology, atmospheric effects, and that of anthropogenic sources. Random oscillatory transients were also noted. Here we will consider the sustained behavior. Fourier transform (FT) analysis spanning ten orders of power and six orders in frequency indicated a background of Gaussian inverse square ($f^{-2}$) non periodic noise power. Which is punctuated by three prominent resonance peaks in the power spectrum marked 1, 2 & 3, respectively as shown in figure 4.

The first and second modes were modeled in a resonant harmonic system and ascribed to the oscillations of the Galilean (~$3\times10^{-1}$ Hz) and torsional (~$8\times10^{-4}$ Hz) modes respectively. These frequencies match the model characteristics of the pendulum when accounting for the length and elastic torsion constant of the suspension fiber. The model pendulum transfer function was fitted to these peak frequencies and also shown in figure 4. At the lowest limit (~$10^{-6}$ Hz), the broad structure in the periodic acceleration of the pendulum is consistent with the frequency of tides. As sun and moon move, their angular positions change in the sky; variation of the horizontal component of the tidal forces with the sun and moon azimuth, $\Omega_{sun}$ and $\Omega_{moon}$ respectively are shown in figure 5, where the angles are plotted from 0° to 7x360° over seven cycles in the extended scheme.

In order to deconvolute the tidal signals, the measured COI time-series was sub-sampled at a rate of 1/60 Hz and then filtered using a low frequency fifth-order, 30 min low-pass Butterworth filter. The phase shift inherent in the application of this type of filter was removed by passing the data through the filter backward and forward. The resultant low passed time-series is shown in figure 6.

In addition, over the period of the experiment, theoretical values of the horizontal tidal acceleration due to Sun and Moon at the location of the laboratory were calculated using the HW95 tidal catalogue [14-16] based on the DES200 JPL Ephemerides. For comparison with the experimental data the theoretical results were projected along the observed axis.

A least-square harmonic procedure was carried out on both the low-passed experimental signal (see figure 7) and the theoretically calculated signal (not shown here). For this procedure the traditional representation of tides was chosen as shown by equation 1 given below:

$$S(t_i)_{horz} = A_0 + \sum_{k=1}^{n} A_k \cdot \cos(\tfrac{2\pi}{T_k}t_i + \phi_k) + R(t_i) \qquad (1)$$

Both the low-passed observed and theoretical time series were fitted to equation 1 using a least squares regression analysis. Specifically a two (N =2) component expansion with periods corresponding to the principal diurnal and principal semidiurnal periods of 23.9345 and 12.4206 hours respectively were fitted; S(t), represents the experimental or theoretical time series while R(t) corresponds to the residual left out after the first and second terms in the RHS of equation 1, is subtracted from S(t). Here $A_0$ is the mean value and the coefficients are the amplitudes and phases of the diurnal ($A_1$, $\phi_1$) and semidiurnal ($A_2$, $\phi_2$) harmonic components corresponding to the principal Solar and Lunar cycles.

The amplitudes and phases derived together with their error at the 95% confidence interval are listed in Tables 1 and 2 below.

**Table 1.** Results of Harmonic Analysis on Experimental Data

|  | Period (hrs) | Amplitude ($10^{-7}$ m·s$^{-2}$) | 95% CI | Phase (degs) | 95% CI |
|---|---|---|---|---|---|
| $A_o$ | - | 2.3855 | 0.0406 | - | - |
| $A_1$ | 23.9345 | 2.7815 | 0.0813 | 91.1327 | 45.0384 |
| $A_2$ | 12.4206 | 4.2466 | 0.0813 | 158.6772 | 44.7548 |

**Table 2.** Results of Harmonic Analysis on Theoretical Estimates (HW95)

|  | Period (hrs) | Amplitude ($10^{-7}$ m·s$^{-2}$) | 95% CI | Phase (degs) | 95% CI |
|---|---|---|---|---|---|
| $A_o$ | - | -2.6596 | 0.1810 | - | - |
| $A_1$ | 23.9345 | 2.8203 | 0.3624 | 93.2541 | 44.9970 |
| $A_2$ | 12.4206 | 4.2381 | 0.3626 | 163.0715 | 44.6773 |

The similarity of the amplitude ratios and phase differences between the experimental data and the theoretical values manifest a correlation between our observations and the theoretical values of the tide (figure 7) confirming that both time-series represent the same (tidal) phenomenon.

In summary, guided by the broad peak structure in the power spectrum supported by comparative harmonic analyses of the observed acceleration and that of the components of the theoretical tidal forces projected along the axis of measurement we conclude that the observed diurnal and semi-diurnal oscillations (~ μHz) are produced by the horizontal components of the solar and lunar tidal (~$10^{-7}$m/s$^2$) accelerations. To the best of our knowledge this is the first reported detection of the horizontal tidal acceleration with a pendulum. Tidal modulations in the vertical free–fall acceleration g is routinely included in high precision force metrology [17-19], our results indicate that in appropriate situations the horizontal components may also have to be accounted for and provide valuable information [20].

**Acknowledgements**

George Voulgaris acknowledges partial support from the National Science Foundation under the Awards: OCE-0451989 and OCE-0535893.

**Figure Captions**

Figure 1**:** Set up to observe the geodesic deviation of a pendulum with respect to the laboratory. The essential parts of the optics and electronics are shown.

Figure 2: Fourier transform power spectrum of the pendulum position signal (red) and its Galilean and tortional model fit (black). The three peaks identified correspond to the relatively high frequency Galilean pendulation (1) torsional oscillation (2), and the very low frequency tidal mode (3).

Figure 3: Four principal modes of oscillations of the cylindrical pendulum. The simple Galilean ($\theta$), torsional ($\phi$), rocking ($\Omega$), and bobbing ($z$) degrees of freedom are illustrated.

Figure 4: Effects of mode–mode coupling. The Y-Z plane elevation view (left panel) shows how see-saw or rocking monition, i.e., changes in $\Omega_*$ causes associated changes in apex angle ($\theta^*$). Even the Galilean mode is not pure, as the X-Z plane, head on elevation illustrates that realistically the system it is more like a double pendulum with two apex angles $\theta_1$ and $\theta_2$.

Figure 5: Changes in the horizontal components of the solar and lunar tidal forces with the sun and moon azimuth, $\Omega_{sun}$ and $\Omega_{moon}$ respectively over seven cycles are shown. The angles are plotted from 0° to 7x360° in the extended scheme.

Figure 6: Time-series of the low frequency, double passed 30 minute Butterworth filtered experimental signal.

Figure 7: Time-series of the theoretical HW95 tidal signal due to Sun and Moon and the harmonic fits (equation 2) to both experimental and HW95 signal.

Figure 1

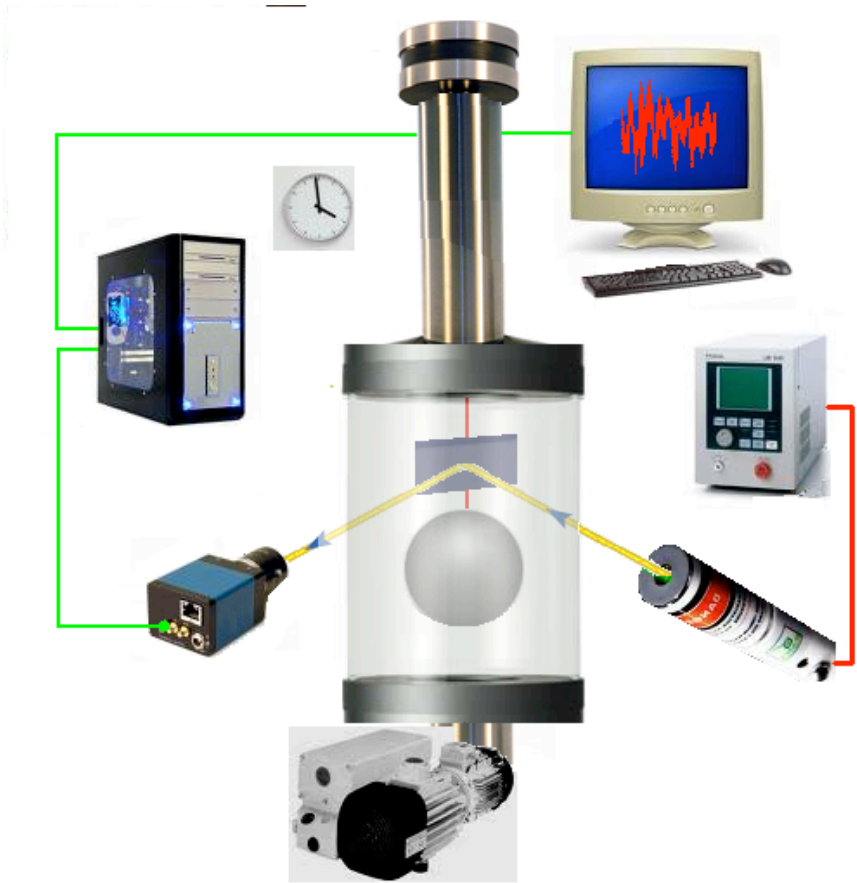

Figure 2

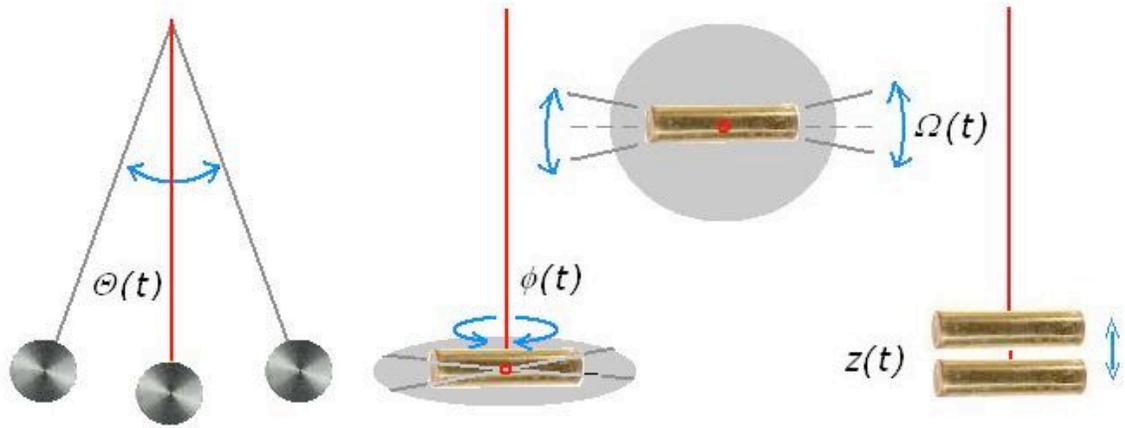

Figure 3

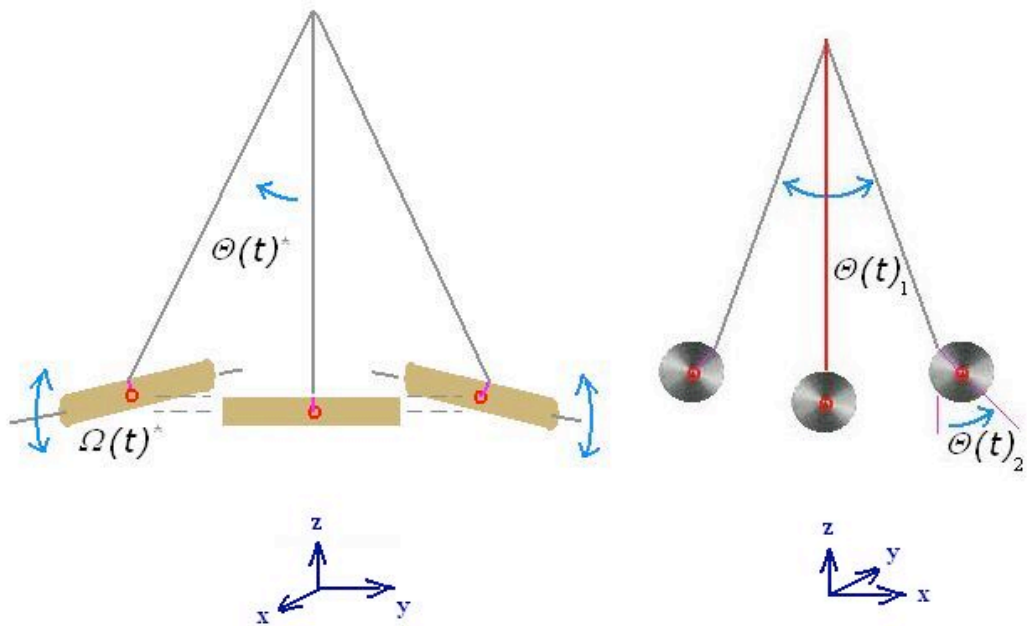

Figure 4

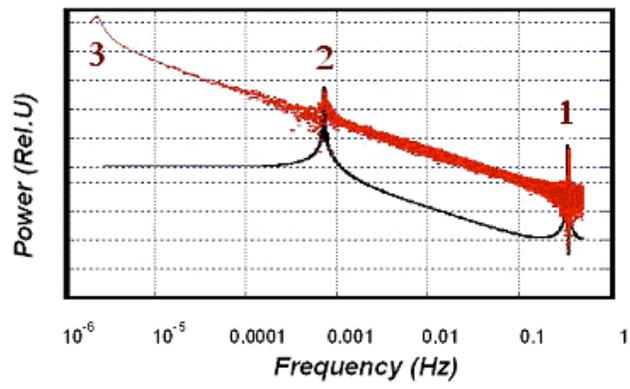

Figure 5

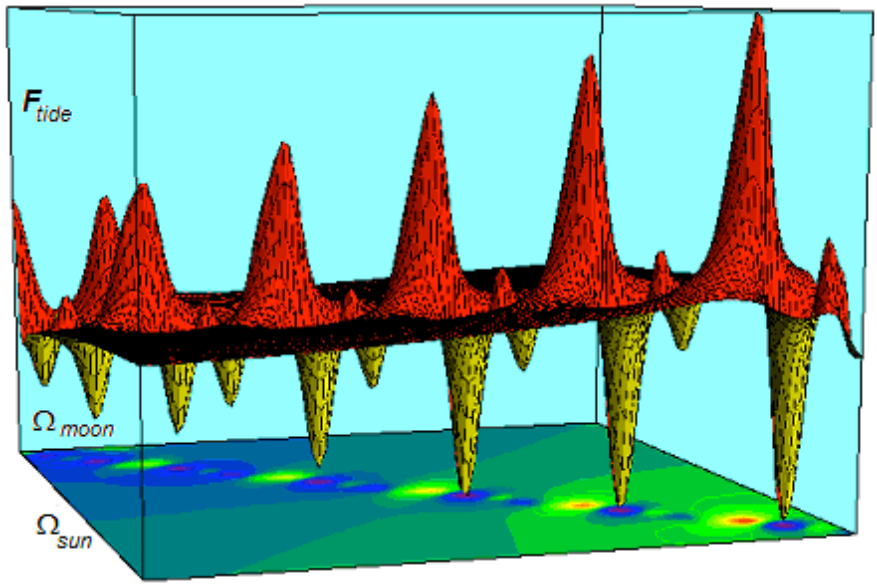

Figure 6

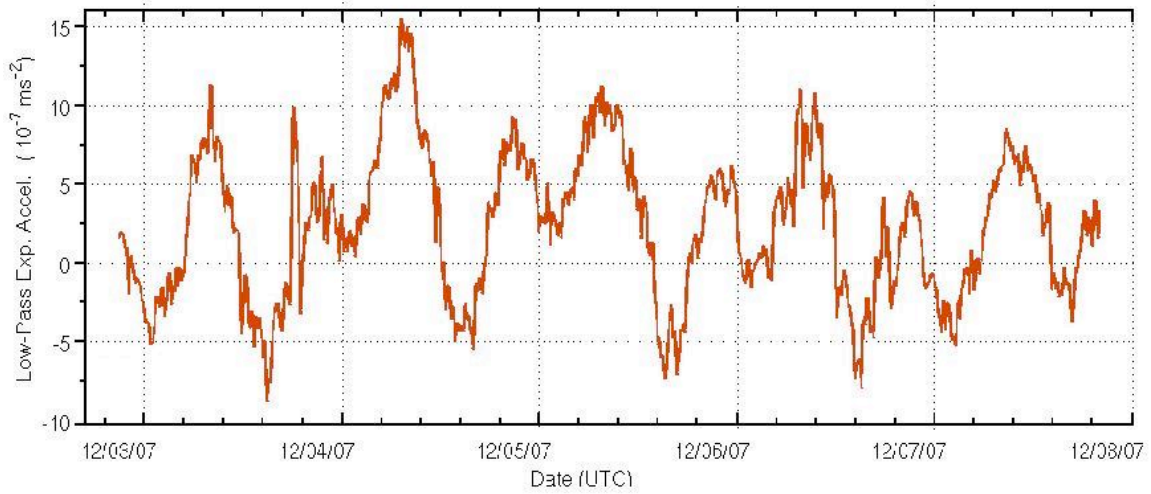

Figure 7

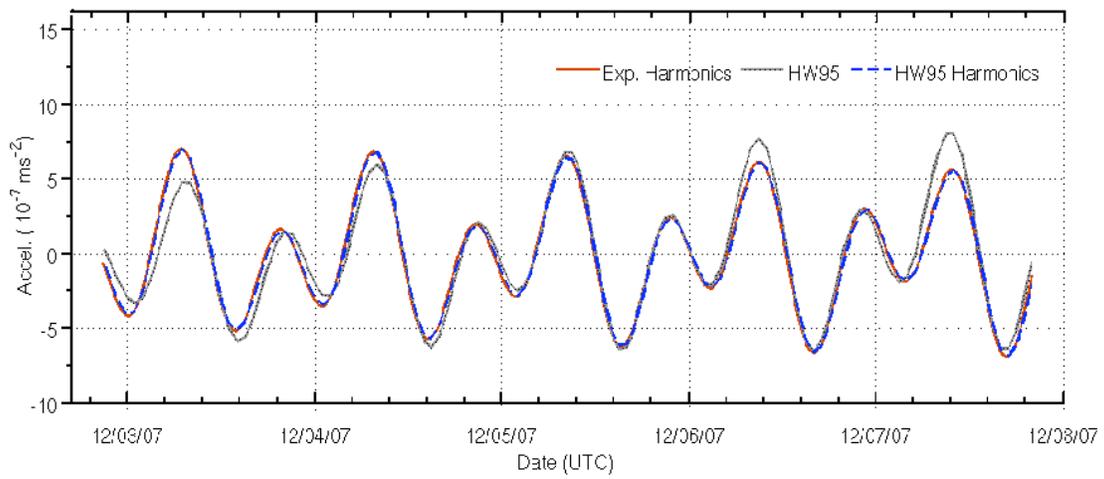